\documentclass[
amsmath,amssymb,
twocolumn,
prb,
superscriptaddress,
floatfix,
longbibliography,
aps, 10pt
]{revtex4-2}

\usepackage{graphicx}
\graphicspath{ {./} }
\usepackage{mathtools}
\usepackage{dcolumn}
\usepackage{bm}
\usepackage{hyperref}
\usepackage{bookmark}
\usepackage[dvipsnames,table]{xcolor}
\usepackage{physics}

\begin{document}
	
\title{Phonon spectrum in the spin-Peierls phase of CuGeO$_3$}

\author{L. Spitz}
\affiliation{Quantum Criticality and Dynamics Group, Paul Scherrer Institute, CH-5232 Villigen-PSI, Switzerland}
\affiliation{Institute for Quantum Electronics, ETH Zurich, CH-8093 H\"onggerberg, Switzerland.}

\author{A. Razpopov}
\affiliation{Institut f\"ur Theoretische Physik, Goethe-Universit\"at Frankfurt, 60438 Frankfurt-am-Main, Germany}

\author{S. Biswas}
\affiliation{Institut f\"ur Theoretische Physik, Goethe-Universit\"at Frankfurt, 60438 Frankfurt-am-Main, Germany}

\author{H. Lane}
\affiliation{Department of Physics and Astronomy, School of Natural Sciences, The University of Manchester, 
Oxford Road, Manchester M13 9PL, UK}
\affiliation{The University of Manchester at Harwell, Diamond Light Source, Didcot, Oxfordshire OX11 0DE, UK}
\affiliation{School of Physics and Astronomy, University of St.~Andrews, St.~Andrews KY16 9SS, UK}

\author{S. E. Nikitin}
\affiliation{PSI Center for Neutron and Muon Sciences, CH-5232 Villigen-PSI, Switzerland}

\author{K. Iida}
\affiliation{Neutron Science and Technology Center, Comprehensive Research Organization for Science and Society (CROSS), Tokai 319-1106, Ibaraki, Japan}

\author{R. Kajimoto}
\affiliation{Materials and Life Science Division, J-PARC Center, Japan Atomic Energy Agency, Tokai, Ibaraki 319-1195, Japan}

\author{M. Fujita}
\affiliation{Institute for Materials Research, Tohoku University, Sendai, 980-8577, Japan}

\author{M. Arai}
\affiliation{European Spallation Source ERIC (ESS), P.O. Box 176, SE, 221 00 Lund, Sweden}

\author{M. Mourigal}
\affiliation{School of Physics, Georgia Institute of Technology, Atlanta, GA 30332, USA}

\author{Ch. R\"uegg}
\affiliation{Quantum Criticality and Dynamics Group, Paul Scherrer Institute, CH-5232 Villigen-PSI, Switzerland}
\affiliation{Institute for Quantum Electronics, ETH Zurich, CH-8093 H\"onggerberg, Switzerland.}
\affiliation{Department of Quantum Matter Physics, University of Geneva, CH-1211 Geneva, Switzerland}
\affiliation{Institute of Physics, Ecole Polytechnique F\'ed\'erale de Lausanne (EPFL), CH-1015 Lausanne, Switzerland}

\author{R. Valentí}
\affiliation{Institut f\"ur Theoretische Physik, Goethe-Universit\"at Frankfurt, 60438 Frankfurt-am-Main, Germany}

\author{B. Normand}
\affiliation{PSI Center for Scientific Computing, Theory and Data, CH-5232 Villigen-PSI, Switzerland}

\begin{abstract}
CuGeO$_3$ has long been studied as a prototypical example of the spin-Peierls transition in a $S = 1/2$ Heisenberg chain. Despite intensive investigation of this quasi-one-dimensional material, systematic measurements and calculations of the phonon excitations in the dimerized phase have not to date been possible, leaving certain aspects of the spin-Peierls phenomenon unresolved. We perform state-of-the-art density functional theory (DFT) calculations to compute the electronic structure and phonon dynamics in the low-temperature dimerized phase. We also perform high-resolution neutron spectroscopy to measure the full phonon spectrum over multiple Brillouin zones. We find excellent agreement between our numerical and experimental results that extend to all measurement temperatures. Notable features of our phonon spectra include a number of steeply dispersive modes, nonmonotonic dispersion features, and specific phonon anticrossings, which we relate to the mode eigenvectors. By calculating the magnetic interactions within DFT and studying the effects of different phonon modes on the superexchange paths, we discuss the possibility of observing spin-phonon hybridization effects in experiments performed both in and out of equilibrium.
\end{abstract}

\maketitle

\section{Introduction}
\label{sintro}

Historically, Peierls' observation concerning a lattice instability in one-dimensional (1D) metals \cite{peierls56} was introduced for magnetic systems in the context of solid ionic free radicals \cite{mccon62,thoma63} and identified in organic materials well described as very weakly interacting $S = 1/2$ chains \cite{jacob76}. Although lattice effects are expected to be stronger in inorganic materials, the spin-Peierls transition was discovered in CuGeO$_3$ at the rather low temperature $T_{SP} = 14.2$~K \cite{hase93}. Extensive experimental efforts quickly characterized the structure \cite{loren94, hirot94,pouge94,hirot95}, thermodynamic properties \cite{sahli94,takah95,liu95,poiri95}, magnetic-field effects \cite{hase93a,hori94}, and spectroscopic properties \cite{udaga94, nishi94,kuroe94,teras95,popov95,loa96}. Definitive measurements of the undimerized and dimerized lattice structures \cite{hirot94,pouge94,brade96} established the basis for further investigation, and these are shown in Fig.~\ref{fig:scheme} together with schematic representations of the (alternating) Cu-O chains and the corresponding magnetic excitation spectra. 

The application of multiple spectroscopic techniques led to increasingly accurate measurements of the phonon modes and magnetoelastic coupling \cite{kuroe94,popov95,brade98a,brade02}. However, several of these early experiments were not verifiably in the low-temperature dimerized phase, and additional features have been discovered in the spectrum when dimerization could be confirmed \cite{damas97,popov98,takeh00}. Definitive measurements of the magnetic excitations over the full Brillouin zone require inelastic neutron scattering (INS), and the magnetic spectrum at zero field and pressure was established in Refs.~\cite{nishi94,regna96a,arai96}. These early studies also raised, and attempted to answer, questions about the nature of the spin-Peierls transition and its accompanying magnetoelastic phenomena \cite{loren94,hirot95,muthukumar1996}.

Theoretical studies of the spin-Peierls transition in a Heisenberg chain \cite{pytte74} were updated by Cross and Fisher \cite{cross79} with the insight that the transition may take place with or without an accompanying soft phonon. With the discovery of CuGeO$_3$, theoretical studies of the appropriate spin Hamiltonian led to a debate between purely 1D character with a strong second-neighbor interaction, $J_2$, introducing Majumdar-Ghosh physics \cite{casti95,riera95,muthukumar1997} and the more 2D character revealed by INS studies of the triplet dispersion \cite{uhrig97a,knett01}. Theoretical work addressing magnetoelastic effects and the nature of the spin-Peierls transition in CuGeO$_3$ \cite{gros98,werne99} reconciled the observed lack of a soft phonon with the conventional theoretical framework \cite{cross79}. The identification of Peierls-active phonons, and of those modes with the strongest spin-phonon coupling coefficients, was pursued in a sequence of studies \cite{brade96,gros98,werne99,feldk02}. 

Early electronic structure calculations based on density functional theory (DFT) found undimerized CuGeO$_3$ to be a metal, with a charge gap opening only due to dimerization and magnetism \cite{matth94,popovic95}. These results highlighted the challenge of dealing with the strongly correlated Cu$^{2+}$ ion, and efforts to capture the insulating gap and the magnetism within DFT were formalized by applying variants of the LDA+U exchange-correlation functional (meaning the local density approximation plus Hubbard-type on-site correlations) \cite{anisi97,dudarev} in a number of studies \cite{sljiv97,wu99,nisik07, filip07,marci21}. However, no DFT method captures all the magnetic energy of the singlet ground state of the Heisenberg dimer, which complicates an accurate analysis of the energetics in the dimerized state of CuGeO$_3$, although cluster calculations have been used to obtain realistic estimates \cite{tohya97}. 

A number of factors make this a good time to revisit the problems posed by CuGeO$_3$. On one hand, the state of the art in both DFT for correlated electronic systems and high-resolution INS for all energies and wave vectors has advanced significantly over the decades. On the other, the advent of ultrafast laser technology capable of both pumping and probing condensed-matter systems at their intrinsic frequencies \cite{zhang17b,salen19} opens the possibility of nonequilibrium methods for exploring dynamical phenomena \cite{torre21}. Intense THz pulses drive the lattice far from equilibrium, displaying nonlinear phononic effects \cite{foers11,subed14,juras17}, and have been used in ordered magnetic materials to manipulate magnetizations \cite{kampf11,mikha15,foers15}. In this context, CuGeO$_3$ presents an excellent candidate for the study of ``magnetophononic'' phenomena \cite{fechn18}, meaning the strong mutual renormalization of the lattice and spin sectors, and the demonstration of phononic effects on the magnetic interactions in a number of materials \cite{disa20,afana21,giorg23} suggests that ultrafast methods can be used to study magnetoelastic dynamics in ways not previously available. The goal of our study is therefore to apply modern methodology to achieve a systematic unification of DFT and INS for the phonon spectrum of CuGeO$_3$, setting the stage for experiments in which coherent phonon driving could be used to address the lattice and spin dynamics both in and out of equilibrium \cite{yarmo23,giorg23}. 

The structure of this article is as follows. In Sec.~\ref{sec:mm} we provide a systematic introduction to the features of CuGeO$_3$ important for our work, to the DFT methodology we apply to compute both the phonon spectrum of the dimerized phase and the magnetic interactions, and to our state-of-the-art INS measurements of the spin and phonon spectra. In Sec.~\ref{sec:comp} we compare our theoretical and experimental results, with particular focus on anomalous dispersive features, on phonon anticrossings, and on the insight gained from inspecting the phonon eigenvectors. In Sec.~\ref{sec:sp} we turn our attention to spin-phonon coupling effects, computing the magnetic interactions within DFT, analyzing the effects of different phonon modes on the superexchange paths, and searching for hybridization features in our spectral data. In Sec.~\ref{sec:dc} we conclude by discussing the context of our results.

\section{Material and methods}
\label{sec:mm}

\subsection{CuGeO$_3$}

\begin{figure}[t]
\includegraphics[width=0.96\linewidth]{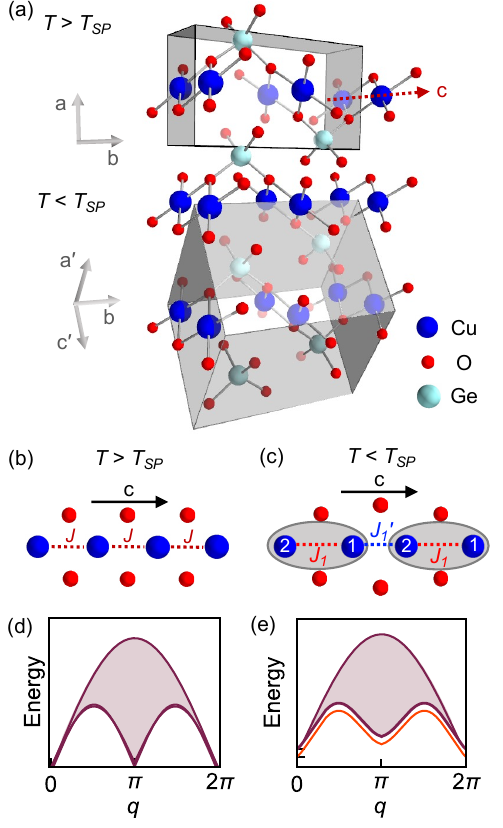}
\caption{(a) Atomic structure of CuGeO$_3$ represented using three planes of CuO$_2$ chains. The upper gray box shows one unit cell (which contains two formula units) of the orthorhombic undimerized structure and the lower gray box one primitive monoclinic unit cell (four formula units) of the dimerized structure. (b) Schematic representation of the uniform (undimerized) spin chain. (c) Representation of the dimerized spin chain, with the two different Cu atoms numbered. The structural dimerization causes the magnetic interaction parameter of the undimerized chain, $J$, to be changed to alternating stronger ($J_1$) and weaker ($J_1^\prime$) interactions. (d) Schematic representation of the magnetic excitation spectrum of the uniform $S = 1/2$ Heisenberg chain, which is a characteristic continuum of two-spinon processes. (e) Excitation spectrum of the alternating chain, which contains a low-energy triplon and at higher energies shows remnants of the two-spinon continuum.}
\label{fig:scheme}
\end{figure}

The primary structural component of CuGeO$_3$ is CuO$_2$ chains, in which the fourfold planar O coordination of the Cu$^{2+}$ ions ensures Cu-O-Cu bond angles close to 90$^\circ$. The quasi-1D magnetic nature of the material is ensured by weak interchain coupling through the Ge atoms. In its undimerized structure, shown at the top of Fig.~\ref{fig:scheme}(a), CuGeO$_3$ is orthorhombic, with space group 51, for which we use the notation $Pbmm$, and room-temperature lattice constants $a = 4.796$ \AA, $b = 8.466$ \AA, and $c = 2.940$ \AA~\cite{brade96}. Because the unit cell contains 10 atoms (two formula units), the undimerized lattice has 30 independent phonon modes.

The dimerized lattice structure involves a doubling of the unit cell not only in the $\hat{c}$ but also in the $\hat{a}$ direction, and has the orthorhombic space group $Bbcm$ (64). In fact this is not the primitive unit cell, which is monoclinic (space group $P2/m$) with lattice constants $a = 5.622$ \AA, $b = 8.402$ \AA, $c = 5.622$ \AA, and angles $\alpha = 90^\circ$, $\beta = 116.84^\circ$, $\gamma = 90^\circ$, and is depicted in the lower half of Fig.~\ref{fig:scheme}(a). In the course of this research we noticed a typographic error in the modern literature for the unit-cell size\cite{rmp}, and we have quoted the low-temperature parameters of Ref.~\cite{brade02}. We did not correct the unit-cell size in our calculations of the phonon spectrum (Sec.~\ref{sec:mm}B), where the effect of the discrepancy is expected to be minimal, but did correct it in our calculations of the superexchange interactions (Secs.~\ref{sec:mm}C and \ref{sec:sp}A). Because the primitive unit cell, with four formula units, has 60 phonon modes rather than the 120 of the orthorhombic unit cell, we use the primitive unit cell in our DFT calculations. For presentation and comparison of the calculated results with the INS data, we use the orthorhombic unit cell of the undimerized phase. 

For a schematic overview of the magnetic properties, Fig.~\ref{fig:scheme}(b) represents the spatially uniform $S = 1/2$ Heisenberg chains aligned along the $\hat{c}$ axis of the undimerized structure and Fig.~\ref{fig:scheme}(d) depicts the magnetic excitation spectrum, which is the des Cloizeaux-Pearson-Faddeev continuum arising from fractionalization of spin-flip processes into fully deconfined spinons (elementary $\Delta S^z = 1/2$ excitations) \cite{Giamarchi2004}. The essence of the spin-Peierls transition is that the magnetic energy gain from dimerizing the spin degrees of freedom into spin singlets on the stronger $J_1$ bonds, represented by the gray ellipses in Fig.~\ref{fig:scheme}(c), more than offsets the energy loss incurred by distorting the lattice structure. The magnetic excitation spectrum is then dominated by a ``triplon'' branch, marked by the orange line in  Fig.~\ref{fig:scheme}(e), which is a gapped $\Delta S^z = 1$ excitation that is threefold degenerate at zero field. However, at higher energies it has been found that the spectral signature of deconfined spinons (pink shading) persists \cite{arai96}. This minimal picture of the spin-Peierls phase would be described by the Heisenberg Hamiltonian
\begin{equation}
H_{\text{s}} = \sum_i J_1 \vec{S}_{1,i} \cdot \vec{S}_{2,i} + J_1^\prime \vec{S}_{2,i} \cdot \vec{S}_{1,i+1}, 
\end{equation}
where $J_1 = J(1 + \delta)$ and $J_1^\prime = J(1 - \delta)$, with $\delta$ a single parameter for the degree of dimerization of the chain interactions. As noted above, early discussions of the magnetic spectrum in CuGeO$_3$ debated whether this minimal model should be augmented by a next-neighbor intrachain coupling ($J_2$), an interchain coupling, or both \cite{casti95,riera95,muthukumar1997,uhrig97a,knett01}. 

\subsection{DFT calculation of the phonon spectrum}

The first step in calculating the phonon spectrum in the spin-Peierls phase of CuGeO$_3$ is to find the structural energy minimum corresponding to the dimerized state within DFT. For this we performed structural optimizations using the VASP code \cite{Kresse_VASP} within the generalized gradient approximation (GGA) \cite{GGA}, in combination with projector augmented-wave (PAW) pseudopotentials \cite{blochl_projector_1994, kresse_ultrasoft_1999}. The plane-wave cut-off was set at 400 eV and an 8$\times$4$\times$8 {\bf k}-mesh was used in the primitive dimerized unit cell; convergence tests were conducted to verify both the cut-off and the {\bf k}-mesh. Here we note that full optimization, meaning that both the unit-cell parameters and the internal atomic positions are allowed to relax, leads to a strong shrinking (by 2.3\%) of the experimentally determined unit cell of the dimerized phase. With the aim of achieving a realistic DFT structure, we therefore followed the constrained relaxation approach, keeping the cell parameters fixed at the experimental values while allowing the internal atomic positions to relax. 

As noted in Sec.~\ref{sintro}, the fact that the strong magnetic correlations in CuGeO$_3$ form a net nonmagnetic (singlet) state poses a major challenge to DFT. To obtain a realistic dimerized structure, we performed the constrained structural relaxation by considering a nonmagnetic configuration of Cu ions. We also included van der Waals corrections \cite{grimme2010consistent,grimme2011effect} to account for possible longer-range interaction effects on the structural optimization, finding that these corrections indeed contribute to a stabilization of the dimerized phase. In this way we obtain a structural dimerization that we quantify from the ratio of the Cu-Cu distances [Fig.~\ref{fig:scheme}(c)] as $d$(Cu$_2$-Cu$_1$)/$d$(Cu$_1$-Cu$_2$) = 1.012 (cf.~the experimental value of 1.008 \cite{brade96}).

By contrast, introducing magnetism in the structural relaxation requires the assumption of a long-range-ordered antiferromagnetic configuration of the Cu atoms, which neglects the quantum magnetic fluctuation effects associated with the formation of singlets. The consequence is that structural dimerization is fully suppressed, because the energy of the ordered magnetic state is optimized by the uniform structure. This competing state does not represent the physics of CuGeO$_3$, and hence we use the results of our nonmagnetic relaxations to establish a dimerized structure by DFT. We comment that the addition of a Hubbard-type correlation in the $d$ orbitals of Cu within the DFT+U framework~\cite{dudarev} had no significant effect on the structures obtained from either magnetic or nonmagnetic relaxation, as both are controlled by different types of physics.

The next step is to compute the phonon spectrum in the dimerized phase by fixing the structure to that obtained from the constrained relaxation. For this we used the Phonopy code \cite{phonopy-phono3py-JPCM,phonopy-phono3py-JPSJ}, which calculates the spectrum within the harmonic and quasi-harmonic approximations by applying the direct force-constant method, where the force constants are obtained from the DFT calculations within GGA. We calculated the phonon dispersions along the path $\Gamma$-M-R-$\Gamma$-X-M with $100$ points per segment, using a supercell of size 2$\times$1$\times$2 orthorhombic dimerized cells.    

The inclusion of magnetism within DFT has a significant effect on the electronic structure, generally providing more realistic results even if the true ground state is nonmagnetic \cite{popovic95,zagou97}, and in this regard our results for structural relaxation in CuGeO$_3$ (above) constitute an important exception. In our phonon calculations, we therefore compare the results obtained by using nonmagnetic and magnetic Cu atoms (a complete list of $\Gamma$-point frequencies is provided in the accompanying data repository \cite{datadump}). The phonon frequencies calculated with magnetic Cu have more accurate values when compared with experiment, as we show in Sec.~\ref{sec:comp}, and the acoustic modes have only minimal negative values, reflecting the stability of the ground-state structure. This leads to the conclusion that the magnetoelastic coupling in CuGeO$_3$, captured in part by including the presence of nonzero Cu magnetic moments in the calculation, plays a crucial role in determining the vibrational modes, as shown also for other spin-Peierls-type systems~\cite{pisani2005,pisani2007}. 

Finally, in order to compare with our experimental measurements, we use the Euphonic code \cite{fair2022euphonic} to compute INS intensities. Euphonic is a Python package that calculates phonon bandstructures and scattered intensities from a force-constant matrix given by the DFT calculations. Because the bandstructure is calculated in the primitive unit cell of the dimerized state, we perform a rotation of the results to the orthorhombic unit cell of the undimerized state for comparison with the INS data.

\subsection{DFT determination of superexchange interactions}

We also performed DFT-based calculations to estimate the magnetic interactions in the dimerized phase of CuGeO$_3$. For this we fixed the structure to that determined by experiment \cite{brade02} and applied a total-energy mapping analysis (TEMA)~\cite{glasbrenner2015,zhang21,razpopov2023}, which is a two-step process. In the first step, we calculate the total energies of multiple different ordered magnetic configurations within DFT+U, using a 1$\times$1$\times$2 supercell of the dimerized structure (in total 32 different magnetic configurations). In the second step, we map these total energies to an assumed spin Hamiltonian with only Heisenberg interactions between many near-neighbor site pairs, as described in Sec.~\ref{sec:sp}A. The total number of Heisenberg interactions ($J_m$) is much smaller than the number of magnetic configurations, and we deduce the actual number ($m_{\rm max}$) of interactions offering the optimal solution to this overconstrained problem. 

All of these DFT calculations were performed within the full-potential local-orbital (FPLO)~\cite{fplo} framework (version 21.00-61) using the GGA~\cite{GGA}, where we included the interaction corrections for the strongly localized Cu 3$d$ electrons within the GGA+U approximation using the atomic limit~\cite{koepe09}. Here $U$ is a free parameter for the effective on-site repulsion that was selected from a range between 6.0 eV and 11.0 eV. The {\bf k}-mesh conversion was verified for each space group and magnetic configuration, and the convergence of our calculations was ensured by requiring the electron density to reach a convergence criterion of $10^{-6}$.

\subsection{Inelastic Neutron Scattering}

Neutron scattering experiments were performed on the time-of-flight spectrometer 4SEASONS at the Materials and Life Science Experimental Facility, located at J-PARC in Tokai, Japan~\cite{kajimoto2011fermi}. Single crystals of CuGeO$_3$ were grown by the travelling-solvent floating-zone method and several crystallites with a total mass of approximately 10 g were coaligned. This sample was oriented with the $(0KL)$ plane of the undimerized orthorhombic unit cell lying in the equatorial plane of the instrument and rotated over an angular range of $120^{\circ}$ in steps of $1^{\circ}$ to map a large part of reciprocal space. Taking advantage of the large out-of-plane coverage of the detector bank on 4SEASONS, we were also able to achieve partial coverage of $H$-direction dispersions. The multirepetition mode of this spectrometer makes it possible to collect data simultaneously with different values of the incident neutron energy \cite{nakamura_first_2009}, which for our study were $E_{\rm i} = 10.9$, 13.8, 18.0, 24.4, 35.0, and 54.4~meV. We used a closed-cycle cryostat with a base temperature of 5~K and measured the full spectrum at the three temperatures $T = 5$, 20, and 110~K. 

\begin{table}[b]
\caption{$\Gamma$-point frequencies, $\omega_i$, of the 60 phonons in the dimerized phase of CuGeO$_3$, calculated by DFT. We separate these phonons by symmetry into infrared- (left) and Raman-active modes (right), and by energy into the seven ranges acoustic (white), low (red and yellow), intermediate (green and blue), and high (orange and purple), whose physical characteristics are described in Sec.~\ref{sec:comp}A.}
\renewcommand{\arraystretch}{1.1}
\begin{tabular}{c@{}c@{}c|c@{}c@{}c}
\hline\hline
$\,$ $i$ $\,$ & $\;$ Symmetry $\;$ & $\,$ $\omega_i$ (THz) $\,$ & $\,$ $i$ & $\;\,$ Symmetry $\;$ & $\;$ $\omega_i$ (THz) \\
\hline
1 & $B_{2u}$ & $-0.08$  & 
$\,$ \cellcolor{red!25}6 & \cellcolor{red!25}$B_{2g}$ & \cellcolor{red!25}$2.81$  \\ 
2 & $B_{1u}$  & $-0.05$ & \cellcolor{red!25}
$\,$ 8 & \cellcolor{red!25}$A_{g}$ & \cellcolor{red!25}$2.94$ \\ 
3 & $B_{3u}$  & $-0.02$ & 
$\,$ \cellcolor{red!25}9 & \cellcolor{red!25}$B_{1g}$ & \cellcolor{red!25}$3.03$ \\ 
\rowcolor{red!25}4 & $B_{2u}$  & $1.59$ &  
$\,$ 10 & $B_{3g}$ & $3.20$ \\ 
\rowcolor{red!25}5 & $B_{3u}$  & $2.81$ &    
$\,$ 11 & $B_{2g}$ & $3.31$ \\ 
\cellcolor{red!25}7 & \cellcolor{red!25} $A_{u}$  & \cellcolor{red!25} $2.93$ &    
$\,$ \cellcolor{yellow!25}14 & \cellcolor{yellow!25} $A_{g}$ & \cellcolor{yellow!25}$5.20$ \\ 
\cellcolor{red!25}12 & \cellcolor{red!25}$B_{3u}$  & \cellcolor{red!25}$3.48$ &    
$\,$ \cellcolor{yellow!25}15 & \cellcolor{yellow!25} $B_{1g}$ & \cellcolor{yellow!25} $5.25$ \\ 
\rowcolor{yellow!25}13 & $B_{1u}$  & $4.89$ &   
$\,$ 17 & $B_{3g}$ & $5.61$ \\ 
\rowcolor{yellow!25}16 & $B_{2u}$  & $5.57$ &   
$\,$ 18 & $B_{2g}$ & $5.75$ \\ 
\rowcolor{yellow!25}21 & $B_{2u}$  & $6.21$ &   
$\,$ 19 & $B_{1g}$ & $6.02$ \\ 
\cellcolor{green!25}22 & \cellcolor{green!25}$B_{1u}$  & \cellcolor{green!25}$7.64$ &   
$\,$ \cellcolor{yellow!25}20 & \cellcolor{yellow!25}$A_{g}$ & \cellcolor{yellow!25}$6.05$ \\ 
\rowcolor{green!25}25 & $B_{2u}$  & $7.79$ &   
$\,$ 23 & $B_{1g}$& $7.65$ \\ 
\rowcolor{green!25}26 & $B_{2u}$  & $8.16$ &   
$\,$ 24 & $B_{2g}$ & $7.78$ \\ 
\rowcolor{green!25}27 & $B_{3u}$  & $8.29$ &  
$\,$ 29 & $A_{g}$ & $8.60$ \\ 
\rowcolor{green!25}28 & $A_{u}$  & $8.51$ &  
$\,$ 31 & $B_{3g}$ & $8.95$ \\ 
\cellcolor{green!25}30 & \cellcolor{green!25}$B_{3u}$  & \cellcolor{green!25}$8.62$ &  
$\,$ \cellcolor{blue!25}33 & \cellcolor{blue!25}$B_{1g}$ & \cellcolor{blue!25}$10.18$ \\ 
\cellcolor{green!25}32 & \cellcolor{green!25}$B_{3u}$  & \cellcolor{green!25}$8.96$ &  
$\,$ \cellcolor{blue!25}34 & \cellcolor{blue!25}$A_{g}$ & \cellcolor{blue!25}$10.56$ \\ 
\rowcolor{blue!25}35 & $B_{2u}$  & $10.64$ &  
$\,$ 36 & $B_{3g}$ & $11.51$ \\ 
\cellcolor{orange!25}39 & \cellcolor{orange!25}$A_{u}$  & \cellcolor{orange!25}$15.10$ &  
$\,$ \cellcolor{blue!25}37 & \cellcolor{blue!25}$B_{1g}$&\cellcolor{blue!25} $11.72$ \\ 
\cellcolor{orange!25}40 & \cellcolor{orange!25}$B_{1u}$  & \cellcolor{orange!25}$15.35$ &  
$\,$ \cellcolor{blue!25}38 & \cellcolor{blue!25}$B_{2g}$ & \cellcolor{blue!25}$12.16$ \\ 
\rowcolor{orange!25}41 & $B_{1u}$  & $15.39$ &  
$\,$ 42 & $B_{2g}$ & $15.51$ \\ 
\rowcolor{orange!25}43 & $A_{u}$  & $15.93$ &  
$\,$ 44 & $B_{3g}$ & $16.36$ \\ 
\rowcolor{orange!25}47 & $B_{3u}$  & $17.25$ &  
$\,$ 45 & $A_{g}$ & $17.02$ \\ 
\rowcolor{orange!25}48 & $A_{u}$  & $17.25$ &   
$\,$ 46 & $B_{3g}$ & $17.09$ \\ 
\rowcolor{purple!25}49 & $B_{3u}$  & $19.70$ &  
$\,$ 51 & $B_{2g}$ & $20.21$ \\ 
\rowcolor{purple!25}50 & $B_{2u}$  & $19.90$ &  
$\,$ 54 & $B_{1g}$ & $21.25$ \\ 
\rowcolor{purple!25}52 & $B_{1u}$  & $20.26$ &  
$\,$ 55 & $A_{g}$ & $21.83$  \\ 
\rowcolor{purple!25}53 & $B_{2u}$  & $20.32$ &  
$\,$ 58 & $B_{1g}$ & $22.95$ \\ 
\rowcolor{purple!25}56 & $B_{3u}$  & $22.37$ &  
$\,$ 59 & $A_{g}$ & $23.09$ \\ 
\rowcolor{purple!25}57 & $B_{2u}$  & $22.40$ &  
$\,$ 60 & $B_{1g}$ & $23.74$ \\ 
\hline\hline
\end{tabular}
\label{tab:w}
\end{table}

In INS, the intensity of the phonon signal increases with the square of the wave vector, $|\mathbf{Q}|^2$, whereas the intensity of the magnetic scattering decreases as $\mathbf{Q}$ increases, following the magnetic form factor (here of Cu$^{2+}$). The data collected with higher $E_{\rm i}$ cover a larger volume of $\mathbf{Q}$ space, and therefore by analyzing the $\mathbf{Q}$-dependence of the signal over multiple Brillouin zones we achieved a reliable separation of the magnetic and phononic contributions. The incident energies we use to access the majority of the phonon spectrum in the data presented here are $E_{\rm i} = 24.4$, 35.0 and 54.4 meV. The corresponding resolutions are, respectively, $1.23$, $1.82$, and $2.91$ meV, and refer to an average linewidth of the incoherent scattering at the elastic line; we quote an average value because the resolution varies across reciprocal space due to the shape of the sample. Data reduction and analysis were performed using the software packages \textsc{Utsusemi}~\cite{inamura2013development} and \textsc{HORACE}~\cite{Horace}. Artifacts appear in the intensity signal when neutrons from two Bragg reflections with out-of-plane components $H$ and $-H$ hit the same detector tube simultaneously, and these were removed where necessary by masking out the affected detector tube.

\begin{figure*}[t]
\includegraphics[width=\linewidth]{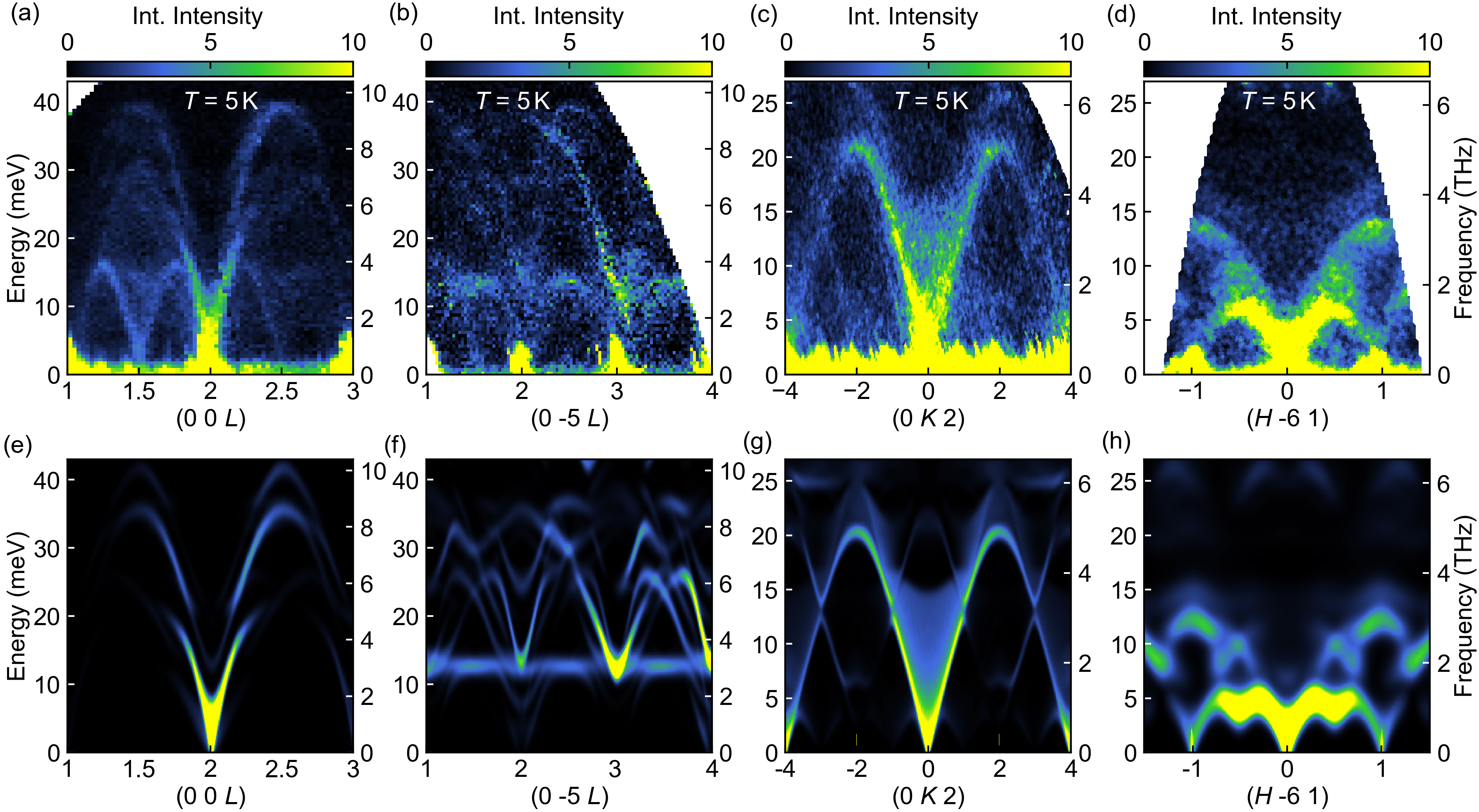}
\caption{Phonon spectra of CuGeO$_3$ at $T = 5$~K measured with $E_{\rm i} = 54.4$~meV (a,b) and $E_{\rm i} = 35.0$~meV (c,d). Upper panels show INS measurements and lower panels DFT calculations. (a) $L$ direction at $H = 0$ with integration ranges $-1 < K < 1$ and $-0.3 < H < 0.3$; the wide integration range in $K$ is justified by the invariance of the dominant phonons in this direction and is chosen to optimize the detector coverage of these branches. (b) $L$ direction at $K = -5$ and $ H = 0$ with integration ranges $-5.1 < K < -4.9$ and $-0.2 < H < 0.2$. (c) $K$ direction at $L = 2$ and $H = 0$ with integration ranges $1.85 < L < 2.15$ and $-0.15 < H < 0.15$. (d) $H$ direction at $L = 1$ and $K = -6$ with integration ranges $-6.15 < K < -5.85$ and $0.85 < L < 1.15$. (e-h) Analogous spectra computed by DFT and displayed with the same integration ranges.}
\label{fig:overview}
\end{figure*}

We will quote all wave vectors in the notation ${\bf Q} = (H~K~L)$, with the indices based on the undimerized unit cell. It is customary in INS to measure excitation energies in meV, but in light-based spectroscopy to measure in THz. We will use both sets of units interchangeably, showing both where it does not complicate the presentation; for the convenience of the reader we state here that the conversion factor is 1 THz $=$ 4.14 meV. The neutron intensities we show throughout the manuscript are not normalized and are compared with the DFT calculations using a single multiplicative factor that is constant throughout the manuscript. All of the wave-vector ranges over which we integrate our intensity data in the two perpendicular directions are consistent between theory and experiment, and have been examined individually to exclude the generation of artifacts. 
 
\begin{figure*}[t]
\includegraphics[width=\linewidth]{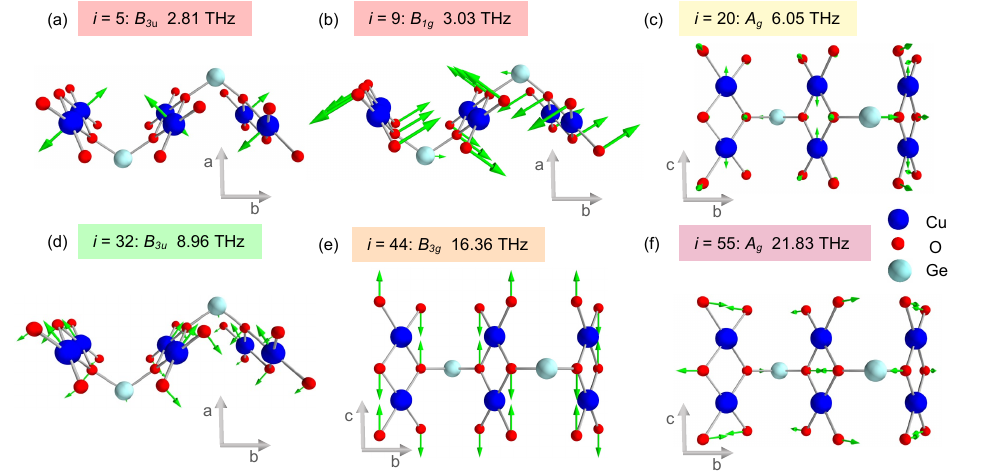}
\caption{Eigenvectors of selected phonons in CuGeO$_3$, representing the atomic displacements in six phonon modes of the dimerized phase. Green arrows represent the relative phases of the atomic motions in each panel and their relative magnitudes across all panels. (a) B$_{3u}$ mode at 2.81 THz. (b) B$_{1g}$ mode at 3.03 THz. (c) A$_{g}$ mode at 6.05 THz. (d) B$_{3u}$ mode at 8.96 THz. (e) B$_{3g}$ mode at 16.36 THz. (f) A$_{g}$ mode at 21.83 THz. }
\label{fig:ev}
\end{figure*}

\section{Phonon properties}
\label{sec:comp}

\subsection{Dispersions}

Table~\ref{tab:w} presents our DFT results for the frequencies of all 60 phonons at the $\Gamma$ point, separated into infrared (IR) and Raman, and color-coded by their energy range. With this basis, we open our discussion of the lattice dynamics by showing in Fig.~\ref{fig:overview} our measured and computed phonon spectra for different paths in reciprocal space. The selected panels include the chain ($L$) direction [Figs.~\ref{fig:overview}(a,b)], the interchain $K$ direction [Fig.~\ref{fig:overview}(c)] in which the magnetic ($bc$) planes lie, and the interplane $H$ direction [Fig.~\ref{fig:overview}(d)]. As a first general remark, we observe a hierarchy of energy scales in the three lattice directions, with the upper edge of the visible spectrum at 40 meV along $L$, 25 meV along $K$, and 15 meV along $H$. Although these are not the highest-energy phonons (Table~\ref{tab:w}), the acoustic-mode velocities reflect a global hierarchy of lattice stiffness coefficients. The steeply dispersing mode visible in Figs.~\ref{fig:overview}(a,c) is the acoustic branch with atomic motions polarized along the chain ($\hat{c}$) direction, and the diffuse intensity visible around $(0\,K\,2)$ in Fig.~\ref{fig:overview}(c) is a consequence of applying the finite integration window in $L$ to this branch. The weakly dispersive mode visible in Fig.~\ref{fig:overview}(d) is caused by the acoustic branches with atomic motions polarized in the inter-chain ($\hat{b}$-axis) and inter-plane directions ($\hat{a}$-axis).

As a second key remark, in Fig.~\ref{fig:overview}(a) we observe rather clearly the magnetic excitation spectrum [shown schematically in Figs.~\ref{fig:scheme}(d,e)], which we will discuss in Sec.~\ref{sec:sp}. However, we note in this context that Fig.~\ref{fig:overview}(b) shows a Brillouin zone that should have no magnetic intensity, in which case we are observing significant phonon intensity in the same parts of reciprocal space, and we examine this overlap in detail in Sec.~\ref{sec:sp}C. Thirdly, we find that our DFT phonon spectra, presented in Figs.~\ref{fig:overview}(e-h), provide excellent quantitative agreement with both the phonon dispersions and intensities, the discrepancy in energy being limited to overestimating the bandwidth by approximately 1-2 meV for the higher-energy modes we observe. Certainly the dominant features of experiment not present in the DFT spectra are only the Bragg peaks and the magnetic scattering in Fig.~\ref{fig:overview}(e), which are in any event not computed. As a technical remark concerning the quality of our DFT calculations and the stability of our dimerized ground state, we draw attention to the extremely small negative energies of the acoustic modes in Table~\ref{tab:w}, a result that has been difficult to achieve in previous studies.

Next we note that, in the spectra of Fig.~\ref{fig:overview}, one does not observe anything close to the 38 phonons of Table \ref{tab:w} with $\omega_i < E_{\rm i}$, but only a single-digit number of the most intense modes. The full phonon dispersions computed by DFT (i.e.~without intensity information) are shown in the accompanying data repository \cite{datadump}. By definition, half of the phonons in Table~\ref{tab:w} are IR-active ($u$) and half are Raman-active ($g$), but this is not a factor in their visibility and  modes of both types are present in Fig.~\ref{fig:overview}. By inspection of Table~\ref{tab:w}, one observes that the phonons at the $\Gamma$ point fall into a number of quite distinct energy ranges, which we have denoted by the color coding. Although these ranges are distinct at $\Gamma$, with a large energy gap above the acoustic phonons followed by significant gaps between the different groups, phonons within one group at the $\Gamma$ point can overlap with other groups on traversing {\bf Q}-space, as Fig.~\ref{fig:overview} makes clear. 

To understand the vibrational properties of all the modes in CuGeO$_3$, we introduce the mode eigenvectors, which are the motional patterns of all the atoms in the unit cell at $\Gamma$. Example eigenvectors are shown in Fig.~\ref{fig:ev} with their phonon index ($i$), symmetry assignment, and frequency color code from Table~\ref{tab:w}. In general, low-frequency modes involve groups of atoms moving together, but in antiphase with other groups, whereas high-frequency modes involve antiphase motions of individual atom pairs with strong covalent bonds providing a large ``spring constant.'' The dimerized phase of CuGeO$_3$ offers alternating counterparts of certain low-frequency ``group-motion'' phonons. With the aid of Fig.~\ref{fig:ev}, we summarize which types of atomic motion have increasing stiffness. 

\begin{figure*}[t]
\includegraphics[width=\linewidth]{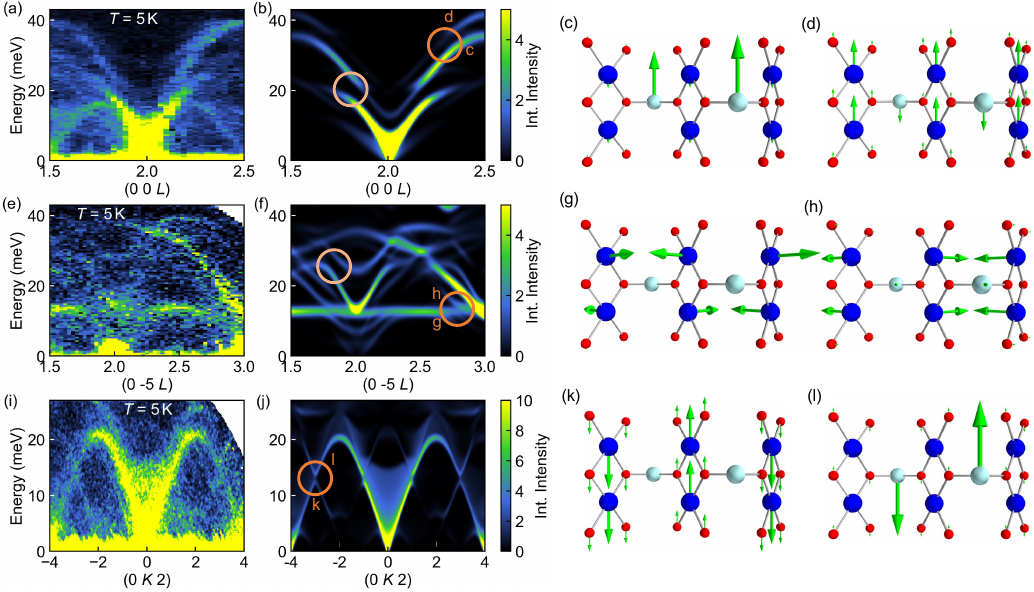}
\caption{Avoided crossings in the phonon spectrum of CuGeO$_3$. INS spectra are shown in the first column and DFT spectra in the second, where orange circles mark prominent phonon anticrossings. The third and fourth columns show the eigenvectors of the two phonon modes involved in the anticrossings marked by the dark orange circles. (a-d) $L$ direction at $H = 0$ with integration ranges $-0.3 < H < 0.3$ and $-1 < K < 1$. (e-h) $L$ direction at $K = -5$ with integration ranges $-0.2 < H < 0.2$ and $-5.1 < K < -4.9$. (i-l) $K$ direction at $L = 2$ and $H = 0$ with integration ranges $-0.15 < H < 0.15$ and $1.85 < L < 2.15$. The incident energies are $E_{\rm i} = 54.4$~meV in panels (a) and (e) and $E_{\rm i} = 35.0$~meV in panel (i).}
\label{fig:ac}
\end{figure*}

\noindent
{\bf White}: acoustic phonons, which reflect different restoring forces for the different lattice directions (Fig.~\ref{fig:overview}).

\noindent
{\bf Red} (1.6--3.5 THz): vibrations out of the buckled $bc$ plane, such as $i = 5$ and 9 in Fig.~\ref{fig:ev}; $i = 5$ [Fig.~\ref{fig:ev}(a)] is the alternating counterpart of a collective Cu-atom motion while $i = 9$ [Fig.~\ref{fig:ev}(b)] is a collective O-atom motion that appears as a rotation of the CuO$_2$ chains.

\noindent
{\bf Yellow} (4.8--6.2 THz): $\hat{a}$-axis motion of all ions and also $\hat{c}$-axis (chain) motion of the Cu ions, as for phonon $i = 20$ in Fig.~\ref{fig:ev}(c), a mode to which we return in Sec.~\ref{sec:sp}B.

\noindent
{\bf Green} (7.6--9.0 THz): several types of Ge-O motion, and also Cu motion within the buckled $bc$ plane, as for phonon $i = 32$ in Fig.~\ref{fig:ev}(d).

\noindent
{\bf Blue} (10.2--12.2 THz): additional motions including global shearing modes of the Cu-O chains (opposing motions of lines of O atoms, as in mode $i = 9$ but directed along the $\hat{c}$-axis). 

\noindent
{\bf Orange} (15.1--17.3 THz): vibrations including antiphase Cu-O motions along $\hat{c}$, as for phonon $i = 44$ in Fig.~\ref{fig:ev}(e).

\noindent
{\bf Violet} (19.7--23.8 THz): vibrations including antiphase Cu-O motions along $\hat{b}$, as for phonon $i = 55$ in Fig.~\ref{fig:ev}(f), where a stiffness even higher than the orange group is caused by O-O restoring forces around the Ge atom.

\noindent
Full eigenvector information and visualization instructions are provided in the accompanying data repository \cite{datadump}. 

In comparison with previous studies, the qualitative features of our measured phonon spectra match well with the systematic measurements and analysis of Ref.~\cite{brade02}, although these were performed for the 30-phonon spectrum of the undimerized phase. Several of the complex anticrossing features we analyze in Sec.~\ref{sec:comp}B can already be found in the undimerized structure. The accuracy and stability of our DFT calculations does not seem to have been achieved in previous studies, which suffered from a combination of unreliable dimerization, unreliable unit-cell parameters, and unreliable stability reflected in some predicted phonon modes having significantly negative energies. In searching for weaknesses of our present DFT calculations, we find mismatches concerning subtle features in the elastically weak $H$ direction, such as the nonmonotonic behavior predicted but not observed around 5 meV ($H = \pm 0.5$) in Figs.~\ref{fig:overview}(d,h), but all other features and energy scales appear to be described very well.  

\begin{figure*}[t]
\includegraphics[width=\linewidth]{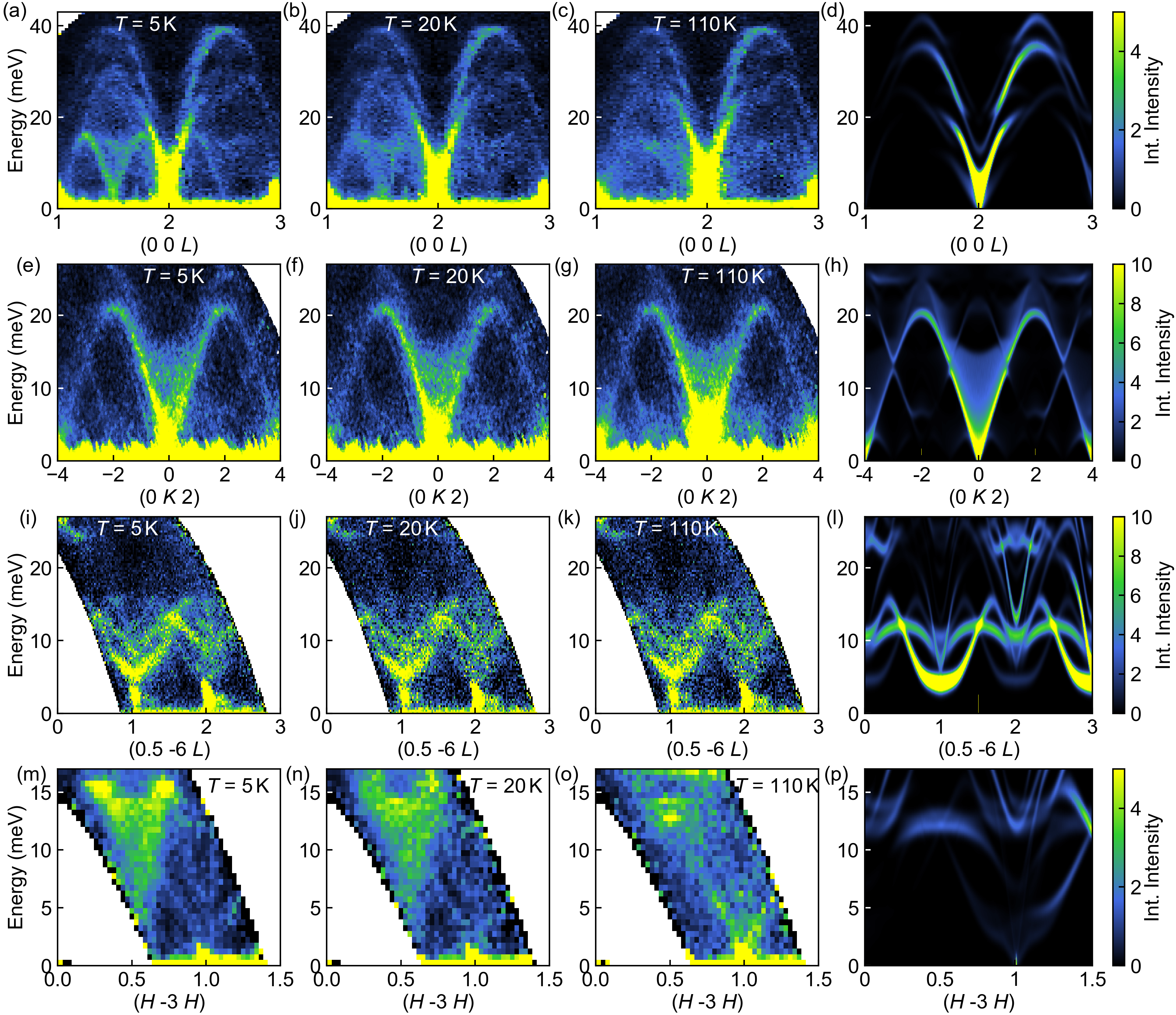}
\caption{Phonon spectra of CuGeO$_3$ measured at 5 K (a,e,i,m), 20 K (b,f,j,n), and 110 K (c,g,k,o) for four selected $(H~K~L)$ paths in the Brillouin zone at incident energies $E_{\rm i} = 54.4$~meV (a-c), $E_{\rm i} = 35.0$~meV (e-g,i-k), and $E_{\rm i} = 24.4$~meV (m-o). Panels (d,h,l,p) show the corresponding DFT calculations. In panels (a-c), (e-g), and (i-k), only minimal thermal effects are discernible, primarily due to the disappearance of the magnetic response in panel (a), whereas in panels (m-o) some more significant changes are evident.}
\label{fig:Tdep}
\end{figure*}

We close this overview by remarking that the $\Gamma$ point of the phonon spectrum (Table~\ref{tab:w}) has special significance for all light-based probes, and also pumps, because for electromagnetic radiation only the very lowest wave vectors have energies matching those of lattice vibrations. In this context, it is important information that the lowest-lying optical phonon is a single mode at 1.6 THz, and that the next candidates lie above 2.8 THz (red group). In a spin-Peierls system such as CuGeO$_3$, one may anticipate experiments both in and out of equilibrium that find a number of low-energy excitations, which one would wish to ascribe with some accuracy to a purely phononic, a purely magnetic, or possibly a mixed magnetoelastic origin.

\subsection{Anticrossings and flat modes}

Phonon modes in the complex spectrum cross each other in wave vector and energy when their symmetries do not match and, barring a cancellation of the coupling, anticross when they do. Phonon anticrossings are quite ubiquitous throughout the spectrum of CuGeO$_3$, and some prominent examples are shown in Fig.~\ref{fig:ac}. Although they must match in energy and symmetry, it is not always true that both modes of an anticrossing should match in intensity, and hence in some cases only one phonon is visible with a gap in its dispersion caused by the near-invisible mode. Once again, the eigenvector information for the anticrossing phonons provides a useful means of interpreting the nature of the mode-mixing, and examples from two different energy scales and three different motional patterns are studied in Fig.~\ref{fig:ac}. 

Although the finite-{\bf Q} eigenvectors can present rather differently from their motional pattern at the $\Gamma$ point (particularly for acoustic phonons), in Figs.~\ref{fig:ac}(a-d) we recognize the $\hat{c}$-axis Ge and Cu motions of our green-band phonons (Table~\ref{tab:w}) and in Figs.~\ref{fig:ac}(e-h) we recognize the out-of-plane Cu motion of certain red-band phonons. A number of similar phonons share the latter type of motion, leading to the appearance of an extended flat band around the maximum of these modes. Figures~\ref{fig:ac}(i-l) introduce a different type of low-energy atomic motion, where the entire CuO$_2$ chains move mostly in antiphase along $\hat{c}$ and are mixed by a $\hat{c}$-axis Ge phonon. 

\subsection{Temperature-dependence}

In our experiments we studied the phonon spectrum at three different temperatures, two spanning $T_{SP}$ and one around the 100~K limit expected for the effects of magnetic fluctuations. In Fig.~\ref{fig:Tdep} we find that none of the phonon dispersions change discernibly, and their scattering intensities at best very minimally, across this full range. This occurs despite the fact that the lattice structure changes from undimerized to dimerized on passing below 14.2~K, and hence one might expect to observe the folding of certain phonon modes between our 20- and 5-K measurements; although the 3 meV splitting reported for one such phonon \cite{popov98} lies close to our energetic resolution at $E_{\rm i} = 54.4$ meV, any associated changes in intensities are difficult to discern in our data. Another consequence of increasing temperature should be a vanishing of the spin response, as we indeed observe in Figs.~\ref{fig:Tdep}(a-c), which could also reveal changes to the phononic contribution. However, the results from our DFT calculations of the phonon spectrum [Figs.~\ref{fig:Tdep}(d,h,l,p)], which we stress were performed for the dimerized structure, continue to provide an excellent account of the observed intensity at all measurement temperatures. From this we obtain an initial suggestion for the possibility we investigate in detail in Sec.~\ref{sec:sp}C, namely that the interplay between the lattice and magnetic spectra can be surprisingly weak even in a spin-Peierls material.

In Fig.~\ref{fig:Tdep} we do observe intensity changes at low energies arising from thermal occupancy (the phonon Bose factor). In Figs.~\ref{fig:Tdep}(m-o) we study a feature reported in Ref.~\cite{brade98a} as a Bose factor and a shift of phonon frequencies. We find that the dominant effect of temperature is the loss of the magnetic signal, which is responsible for the two strong peaks at 16 meV (at $L = 0.3$ and 0.7) in Fig.~\ref{fig:Tdep}(m). The changing magnetic signal improves the visibility of the phonon mode at $L = 0.5$, but we do not have the contrast to discern whether the phonon energies in the 12-15 meV range have changed with temperature [Figs.~\ref{fig:Tdep}(n,o)]. The issue of distinguishing between phononic and magnetic spectral contributions at exactly the same places in wave vector and energy is one to which we return in Sec.~\ref{sec:sp}C.

\section{Spin-phonon hybridization} 
\label{sec:sp}

Although our INS and DFT studies indicate that the dimerization transition has a very limited influence on the phonon spectrum, the lattice structure is fundamental in determining the magnetic interactions. In a $S = 1/2$ magnetic insulator such as CuGeO$_3$, the dominant mechanism for spin interactions is superexchange, which is highly sensitive to the geometry of the interatomic pathway involved. Here we investigate the resulting interplay between structure and magnetism with a view to detecting hybridization effects between the magnetic and phononic spectra, and to observing magnetoelastic phenomena.

\begin{figure*}[t]
\includegraphics[width=\linewidth]{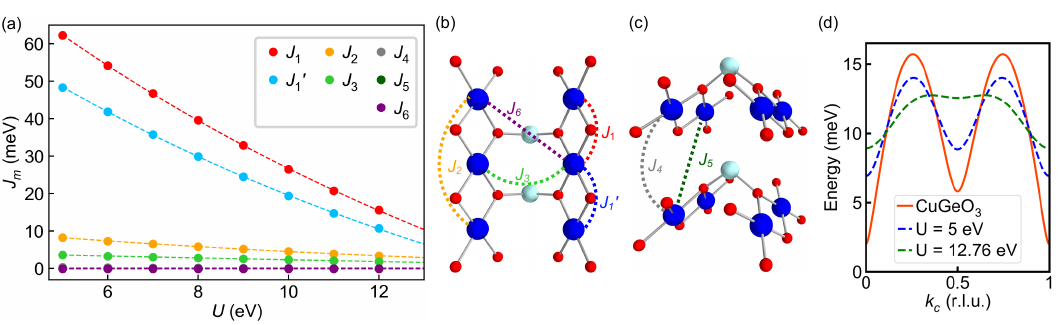}
\caption{Superexchange interactions, $J_m$, calculated by DFT+TEMA using different assumed magnetically ordered configurations. (a) Evolution of superexchange values with the effective on-site Hubbard interaction, $U$, used to compute the DFT energies. (b,c) Representation of the seven near-neighbor Cu-Cu interactions to which the ground-state energies were mapped. (d) Comparison between the triplon dispersion relation of CuGeO$_3$ as a function of $k_c$ (red) and dispersions obtained from the TEMA interaction parameters of panel (a) at $U = 12.76$ eV (blue) and at $U = 5.00$ eV with an additional free scale factor applied to $J_1$ (green).}
\label{fig:J_DFT}
\end{figure*}

\subsection{DFT calculations of superexchange}

We begin by applying the DFT+TEMA procedure~\cite{glasbrenner2015,zhang21,razpopov2023} to estimate the magnetic interaction parameters of CuGeO$_3$. As described in Sec.~\ref{sec:mm}C, the first step of the process is to compute the total energies for the dimerized structure of CuGeO$_3$ in multiple different assumed magnetic configurations. These energies are computed for a broad range of values of the bare repulsion parameter, $U$, between electrons on the same site. The second step is to assume that all the energy differences arise from magnetic interactions, and hence to map these to a Heisenberg Hamiltonian,
\begin{equation}
    H = \sum_{i,j} J_{ij} \vec{S}_i \cdot  \vec{S}_j,
\end{equation}
where the sums run over all site pairs. The Heisenberg Hamiltonian is expected to provide an excellent description of the magnetic sector in Cu$^{2+}$ systems, where the orbital moment is completely quenched, leaving as the only unknown the range of the relevant $m$th-nearest-neighbor superexchange interactions ($J_{ij} \equiv J_m$). We treat the maximum range ($m_{\rm max}$) as a free parameter, which remains much smaller than the number of magnetic configurations we compute, to determine the range of interactions that offers a complete description of the total-energy data. Finally, $U$ is refined by comparison with the magnetic interaction parameters deduced from experiment.

Preliminary TEMA calculations for the undimerized phase indicated that a good description of the total-energy data is obtained by using interactions on the first six near-neighbor bonds ($m_{\rm max} = 6$), whose Cu-Cu distances at the lowest relevant temperature we list for reference in Table \ref{tab:geometry}. For the dimerized phase, our detailed calculations confirmed this result, but with alternating values of the dominant $J_1$ bond distance (Table \ref{tab:geometry}), leading to a bilinear Heisenberg Hamiltonian containing seven superexchange interactions. The values of these interactions computed from the total-energy differences are shown in Fig.~\ref{fig:J_DFT}(a) and their geometries are represented in Figs.~\ref{fig:J_DFT}(b,c). 

\begin{table}[b]
\centering
\renewcommand{\arraystretch}{1.1}
\begin{tabular}{cc|cc}
\hline\hline
\multicolumn{1}{c|}{$\; J_m \;$}  & $\,$ $d$(Cu-Cu) [\AA] $\,$ & \multicolumn{1}{c|}{$\; J_m \;$}  & $\,$ $d$(Cu-Cu) [\AA] $\,$ \\ \hline
\multicolumn{2}{c|}{undimerized}  &  \multicolumn{2}{c}{dimerized}  \\ \hline
\multicolumn{1}{c|}{$J_1$}        &         2.94447     &  \multicolumn{1}{c|}{$J_1$}        &         2.95581          \\
\multicolumn{1}{c|}{} & & \multicolumn{1}{c|}{$J_1^\prime$} &         2.93319         \\
\multicolumn{1}{c|}{$J_3$}        &         4.20128     &  \multicolumn{1}{c|}{$J_3$}        &         4.20090          \\
\multicolumn{1}{c|}{$J_4$}        &         4.78928     &  \multicolumn{1}{c|}{$J_4$}        &         4.78941          \\
\multicolumn{1}{c|}{$J_6$}        &         5.13037     &  \multicolumn{1}{c|}{$J_6$}        &         5.13007          \\
\multicolumn{1}{c|}{$J_5$}        &         5.62202     &  \multicolumn{1}{c|}{$J_5$}        &         5.62214          \\
\multicolumn{1}{c|}{$J_2$}        &         5.88894     &  \multicolumn{1}{c|}{$J_2$}        &         5.88900         \\ 
\hline\hline
\end{tabular}
\caption{Bonds with significant superexchange interactions as revealed by the TEMA analysis, numbered by their computed strengths but ordered by the interatomic Cu-Cu separations measured in experiment at 20 K in the undimerized phase \cite{brade96} and at 4 K in the dimerized phase \cite{brade96}.}
\label{tab:geometry}
\end{table}

In Fig.~\ref{fig:J_DFT}(a) we observe that the $J_1$ and $J_1^\prime$ interactions largely track each other over the full range of $U$ values, with the degree of dimerization rising from around $\delta = 0.1$ at small $U$ to $\delta = 0.2$ at large $U$. Because $J_1$ sets the energy scale of the dispersion, one observes already that the relevant $U$ values are those at the upper end of our test range. All of the longer-ranged couplings can be expected to change very little between the undimerized and dimerized phases, given that their distances (Table~\ref{tab:geometry}) and geometries change only minimally. The next-strongest bond, $J_2$, is in fact the next-neighbor interaction in the chain direction, which was discussed in detail in early CuGeO$_3$ studies \cite{casti95,riera95,muthukumar1996,muthukumar1997,uhrig97a}; the values of $0.15$$-$$0.25 J_1$ we find are rather lower than the proposed fits to a 1D model, but fully consistent with the optimized 2D fits of Ref.~\cite{knett01}. From our calculations, $J_3$ is the only relevant interchain coupling, which will therefore govern the estimated magnetic dispersion in the $\hat{b}$ direction. Finally, we find $J_4$ and $J_6$ to be ferromagnetic, and $J_5$ to change sign as the value of $U$ is changed, but all three are more than two orders of magnitude smaller than $J_1$ [Fig.~\ref{fig:J_DFT}(a)] and hence can be neglected in a discussion of the spin spectrum. 

In general, the $U$-dependence of the antiferromagnetic $J_m$ parameters we extract follows the $1/U$ form anticipated from the simplest expression of superexchange [Fig.~\ref{fig:J_DFT}(a)], although the relative couplings evolve systematically across our $U$ range. In any fit to experiment, the values $J_m$ depend on the theoretical procedure employed, because different procedures capture quantum fluctuation effects at different levels, and in this sense it is not clear which effects are captured by our TEMA parameters. Based on the fact that we find three significant intra-chain $J_m$ parameters, but only one interchain parameter, we consider the simplified triplon dispersion \cite{uhrig97a,demaz25}
\begin{align} \label{eq:disp_2d}
\omega(k_{c}, k_{b}) & \! = \! J_1 \sqrt{1 \! - \! \lambda \cos(k_{c}) \! - \! 2 \mu \cos(\tfrac{1}{2}k_{c}) \cos(\tfrac{1}{2}k_{b})},
\end{align}
where $k_c$ is the wave vector in the chain direction and $k_b$ the wave vector in the orthogonal direction with finite coupling \cite{regna96a}. Here $J_1$ sets the energy scale, $\lambda = (J_1^\prime - 2 J_2)/J_1$ is the effective interdimer propagation in the chain direction, and $\mu = J_3/J_1$ is the interchain term. We seek the best fit by parameterizing $J_1$ and the three leading coupling ratios with $U$ and making a least-squares fit to the full triplon dispersion relation in the 2D $(k_b,k_c)$ space. We comment again that this $U$ is a parameter required within DFT when the LDA+U or GGA+U exchange correlation functional is used, and its precise value depends on the implementation and double-counting considered. Estimates of the actual $U$ for Cu 3$d$ electrons can be obtained within DFT from methods such as constrained RPA \cite{aryasetiawan2004,honerkamp2018}, with values usually falling in the range 4-8 eV, although some estimates vary up to 10 eV~\cite{tesch2022}.

In Fig.~\ref{fig:J_DFT}(d) we show two different fitting approaches, which make clear that the interaction parameters we obtain from the DFT+TEMA procedure cannot capture the very small gap-to-bandwidth ratio of CuGeO$_3$. Extrapolating slightly beyond the largest $U$ values in our calculation achieves the correct overall energy scale ($J_1 = 11.60$ meV at $U = 12.76$ eV), but makes the gap significantly too large, which is mostly a consequence of the small $J_1^\prime$ within the effective dimerization [$\lambda$ in Eq.~\eqref{eq:disp_2d}]. The interchain coupling, which determines the separation of the gap from the saddle point [visible respectively at the edges and center of the red line in Fig.~\ref{fig:J_DFT}(d), i.e.~at 2 and 6 meV], is also somewhat overestimated. In the second fit, we seek the best parameter ratios by allowing $J_1$ to be renormalized freely, and find the best gap-to-bandwidth ratio at our smallest $U$ values, although the required energetic renormalization is very high (a factor of 5.5, as $J_1 = 62.2$ meV at $U = 5.0$ eV). These results emphasize the delicate nature of the triplon dispersion in dimerized CuGeO$_3$, with the very small relative gap reflecting the proximity to the gapless spinon spectrum of the undimerized phase [Fig.~\ref{fig:scheme}(d)]. Given the accuracy of our DFT+TEMA procedure, we do not attempt to apply it for the estimation of the magnetic interactions for specific frozen-phonon configurations, although this would in principle provide a means of estimating the spin-phonon couplings, $g_i$, for each phonon $i$.

\subsection{Changes to superexchange paths}

\begin{figure}[t]
\includegraphics[width=\linewidth]{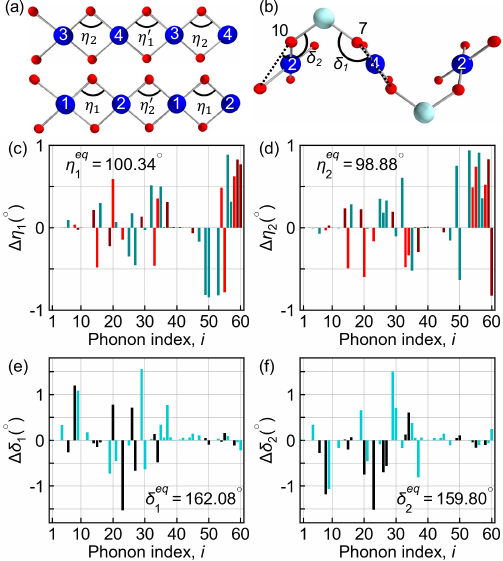}
\caption{Effect of phonon modes on superexchange angles in CuGeO$_3$. (a) Definition of the angles $\eta_1$, $\eta_2$, $\eta_1^\prime$, and $\eta_2^\prime$ parameterizing the geometry of the CuO$_2$ chains, shown for two adjacent chains in the $bc$ plane; numbers indicate the four inequivalent Cu sites in the dimerized unit cell. (b) Definition of the angles $\delta_1$ and $\delta_2$ parameterizing the geometry of the interchain superexchange path through the Ge atoms. (c,d) Changes in the angles $\eta_1$ and $\eta_2$ shown for all the phonons indexed in Table \ref{tab:w}. Green bars mark phonons that have similar effects on $\eta_1$ and $\eta_2$, and hence distort the Cu-O plaquettes. Red bars mark phonons that have opposing effects on $\eta_1$ and $\eta_2$, and hence impact the dimerization. (e,f) Changes in the angles $\delta_1$ and $\delta_2$ shown for all phonons. Turquoise bars mark phonons for which $\delta_1$ and $\delta_2$ change in phase between nearest-neighbor Cu sites along the chains, black bars mark those for which the change is out of phase. All angle changes are calculated from their equilibrium values in the dimerized unit cell, modulated by a single eigenvector with a constant maximum amplitude.}
\label{fig:effect}
\end{figure}

\begin{figure}[btp]
\includegraphics[width=\linewidth]{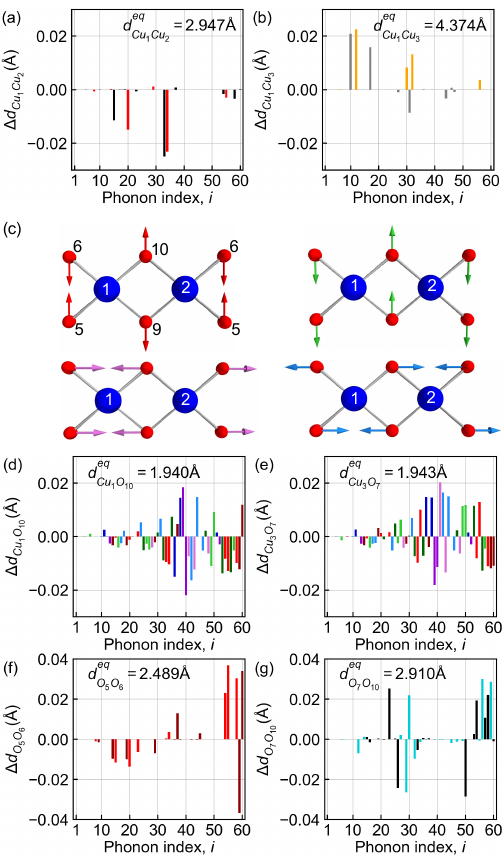}
\caption{Changes in interatomic separations caused by phonon modes in CuGeO$_3$. (a) Intrachain Cu-Cu distance; black bars mark phonons causing the same displacement in neighboring chains, red bars phonons causing opposite displacements. (b) Interchain Cu-Cu distance; orange bars mark phonons causing the same effect on neighboring pairs along the chain direction, grey bars phonons causing opposite effects. (c) Representation of different types of out-of-phase O-atom motion changing the Cu-O distances. (d,e) Cu-O distances in two neighboring chains; O$_7$ and O$_{10}$ are linked to the same Ge atom, as shown in Fig.~\ref{fig:effect}(b). The phonon type is displayed using the colors in panel (c), lighter colors denoting modes with antiphase motions on neighboring plaquettes in the same chain (c), darker colors modes with in-phase motions [obtained from panel (c) by inverting the direction of the center pair of arrows]. (f) Intrachain O-O distance; red bars mark phonons causing opposite displacements on neighboring plaquettes along the same chain, dark red bars phonons with the same effect. (g) Interchain O-O distance for two O atoms neighboring the same Ge atom; black bars mark phonons causing opposite displacements on neighboring sites along the chain direction, turquoise bars phonons causing in-phase displacements.}
\label{fig:effect_dist}
\end{figure}

Much less computationally intensive is to use our DFT results for the phonon spectrum and eigenvectors to investigate the effect of specific phonon modes on the leading superexchange pathways. Although we cannot compute an absolute magnitude for the modulation of each $J_m$ in this way, it is in principle possible to gauge relative effects. Quite generally, the overlap integrals of orbitals forming the superexchange pathway depend on the interatomic separation of each atom pair and on the angles made by each of these pathway segments. In Fig.~\ref{fig:effect} we study the changes away from their equilibrium values of the key bond angles due to each of the 60 $\Gamma$-point phonon modes in Table \ref{tab:w} and in Fig.~\ref{fig:effect_dist} the changes in bond lengths. 

\begin{figure*}[t]
\includegraphics[width=\linewidth]{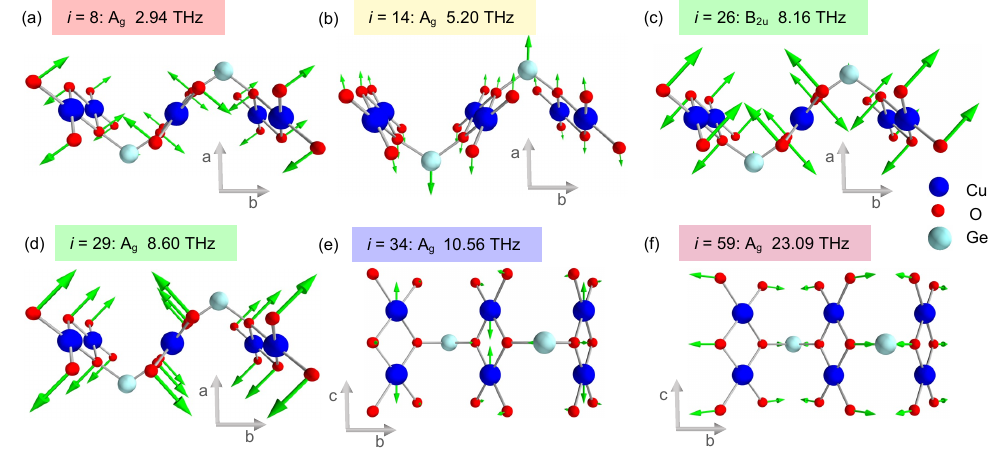}
\caption{Eigenvectors of selected phonons with significant effects on the superexchange angles in CuGeO$_3$. (a) Phonon $i = 8$ has a small effect on the angles $\eta_1$ and $\eta_2$ affecting the intra-chain coupling and a strong effect on the angles $\delta_1$ and $\delta_2$ affecting the interchain coupling. (b) Phonon $i = 14$ has a moderate effect on $\eta_1$ and $\eta_2$ and a small effect on $\delta_1$ and $\delta_2$. (c,d) Phonons $i = 26$ and 29 have moderate effects on $\eta_1$ and $\eta_2$ and strong effects on $\delta_1$ and $\delta_2$. (e,f) Phonons $i = 34$ and 59 have strong effects on $\eta_1$ and $\eta_2$, but only mode 34 has a moderate effect on $\delta_1$ and $\delta_2$. Significant effects of these phonons on the bond lengths are noted in the text.}
\label{fig:effect_vectors}
\end{figure*}

Following the notation of previous studies \cite{brade96,gros98,werne99,feldk02}, in Fig.~\ref{fig:effect}(a) we define the key angular variables $\eta_1^{(\prime)}$ and $\eta_2^{(\prime)}$ describing the geometry of the chains ($\hat{c}$-axis superexchange) and in Fig.~\ref{fig:effect}(b) the angles $\delta_1$ and $\delta_2$ defining the interchain geometry ($\hat{b}$-axis superexchange). At equilibrium, $\eta_1 = \eta_1^\prime$ and $\eta_2 = \eta_2^\prime$, and this relationship is maintained by the symmetry of the majority of the phonon modes in CuGeO$_3$, but not in all. Figures \ref{fig:effect}(c-f) show the relative effects of each phonon on the $\eta$ and $\delta$ angles. Before interpreting these results, we remark that the four phonon modes termed ``Peierls-active'' in Ref.~\cite{werne99} and labelled as $\Omega_j$ with $j = 1$, 2, 3, 4 are our phonons $i = 8$, 20, 34, and 55 in Table \ref{tab:w}, all of which are Raman-active. These modes were identified by their eigenvectors, rather than by their energies or energy sequence, as we find a small discrepancy between the theoretical and experimental mode frequencies that grows steadily with $i$. The eigenvectors of phonons $i = 20$ and 55 are illustrated in Fig.~\ref{fig:ev} and those of phonons $i = 8$ and 34 in Fig.~\ref{fig:effect_vectors}. 

To read the phonon effects on the $\eta$ angles, we focus first on the color of each bar, which codes the intrachain modulation. Light red bars in Figs.~\ref{fig:effect}(c,d) mark phonons whose angle changes cause $J_1$ to increase while $J_1^\prime$ decreases, and hence modulate the chain dimerization, while dark red modes change $J_1$ and $J_1^\prime$ in the same direction. Green bars mark further phonons that also have strong effects on the bond angles, but whose eigenvectors indicate that their primary effect is to distort the Cu-O$_2$-Cu squares in a way that does not necessarily cause a strong change in $J_1$ and $J_1^\prime$. Next we read the relationship between the distortions in adjacent chains by comparing Figs.~\ref{fig:effect}(c) and Figs.~\ref{fig:effect}(d). For the light red modes, opposite changes of $\eta_1$ and $\eta_2$ indicate a global dimerization, from which we observe that phonons 20, 34, and 55 are indeed the relevant modes, whereas phonon 8 has at best a moderate effect and phonons 15, 23, 33, 54, and 58 have a strong effect but the wrong symmetry for a global dimerization. In this regard, the latter are in the same category as the dark red phonons 14, 29, 59 (all shown in Fig.~\ref{fig:effect_vectors}), 37, and 60, while green phonons 32, 35, 49, 50, 53, and 56 also cause very strong effects on the $\eta$ angles. 

Turning to Figs.~\ref{fig:effect}(e,f), here the black bars mark phonons in which the changes in $\delta_1$ and $\delta_2$ alternate along the chain direction, while turquoise bars mark other mode types. First we observe that no phonons beyond $i = 37$ have significant effects on the interchain bond angles, reflecting the very local nature of the high-frequency modes. Phonons $i = 8$, 20, and 34 are among those with the strongest effects on these bond angles, along with 9, 19, 26 [shown in Fig.~\ref{fig:effect_vectors}(c)], 27, 30, and 37, while mode 55 is not. However, the very strongest effects are those of phonons 23 and 29 [shown in Fig.~\ref{fig:effect_vectors}(d)], which are also the only modes (other than 27) in which $\delta_1$ and $\delta_2$ change in the same direction, suggesting a very strong modulation of the interchain interaction $J_3$. 

Turning next to the bond lengths, in Fig.~\ref{fig:effect_dist} we show the changes in Cu-Cu, Cu-O, and O-O separations caused by each of the phonon modes. The first observation in Figs.~\ref{fig:effect_dist}(a,b) is that surprisingly few phonons have a significant effect on the Cu-Cu distances, indicating that most involve collective atomic motions of correlated units larger than a single Cu-Cu spacing (for low $i$) or (for high $i$) Cu-O motions whose effects on this spacing cancel. Modes with a large effect on the intrachain Cu-Cu spacing include 20 [Fig.~\ref{fig:ev}(c)] and 34 [Fig.~\ref{fig:effect_vectors}(e)], as well as 15 and 33 [Fig.~\ref{fig:effect_dist}(a)], while only a limited number of slow modes affect the interchain Cu-Cu spacing [Fig.~\ref{fig:effect_dist}(b)].

To present all of the information about changes in Cu-O bond lengths in a limited number of figures, we follow the approach adopted in Fig.~\ref{fig:effect} and classify the intrachain modulations of the CuO$_2$ plaquettes by using four different colors to indicate different categories of relative O motion. First, we note that all the motional patterns shown in Fig.~\ref{fig:effect_dist}(c) are out of phase along the chain and have in-phase counterparts obtained by inverting the directions of every second pair of arrows, which are represented by the darker versions of each color in Figs.~\ref{fig:effect_dist}(d,e). Second, we caution that this color code is completely independent of the colors used to denote phonon energy ranges in Table \ref{tab:w}. Third, the information about the interchain relationship between all of these changes is obtained by comparing the directions of the colored bars between Figs.~\ref{fig:effect_dist}(d) and \ref{fig:effect_dist}(e). 

In both panels we observe that almost all phonon modes, other than some of the slowest, act to change the Cu-O distances, with a somewhat systematic dependence of the amplitude on $i$ that peaks around $i = 40$ before falling slightly towards the fastest phonons. All of phonons 36-44 are extremely active by the measure of Cu-O bond-length modulation, and all fall in the blue and purple motional categories on the right side of Fig.~\ref{fig:effect_dist}(c) [mode 44 is shown in Fig.~\ref{fig:ev}(e)]. By contrast, the red and green categories are strongly represented among the highest $i$ values, including modes 55 [Fig.~\ref{fig:ev}(f)] and 59 [Fig.~\ref{fig:effect_vectors}(f)] studied above, which we already stated to be mostly local Cu-O vibrations. These rather large effects make clear that many phonon modes can act to alter the superexchange pathways, although strong effects on the Cu-O distances do not correspond directly to strong effects on the superexchange angles, $\eta_{1,2}$, shown in Figs.~\ref{fig:effect}(c,d). In this regard, the red and green motional patterns of Fig.~\ref{fig:effect_dist}(c) do cause large changes in $\eta_{1,2}$ whereas the purple and blue patterns do not. These results show that only some of the strongly active phonons can modulate the $J$ values, while others cancel in their effects, and a still smaller number of these modes cause a significant modulation of particular $J$ ratios, the primary example being that modulation of the dimerization ratio, $J_1^\prime/J_1$, is largely restricted to modes color-coded in light red. 

Finally, intrachain O-O spacings are altered strongly only by the fastest phonons [Fig.~\ref{fig:effect_dist}(f)], including 55 and 59 again, and we deduce that otherwise these O atoms move largely in-phase. The phonons altering interchain O-O separations are not familiar from any of the other measures, and include some modes in the 20s and 50s [Fig.~\ref{fig:effect_dist}(g)]. Although the information contained in the bond distances (Fig.~\ref{fig:effect_dist}) is not fully independent of that in the bond angles (Fig.~\ref{fig:effect}), it is clear that there are no simple relations and hence redundancies.  

To summarize, we have shown that many of the phonon modes in CuGeO$_3$ can have a significant influence on the superexchange pathways within the spin chains, providing a better understanding of their possible hybridization with the magnetic excitations. However, only a small number of phonons have both strong effects and the symmetry of the global dimerization pattern, making them candidates for influencing the spin-Peierls transition. Although the phonon eigenvectors are depicted in Figs.~\ref{fig:ev} and \ref{fig:effect_vectors} only for the $\Gamma$ point, this information delivers significant insight into the origin of modulation effects on the spin interactions that could be used in a nonequilibrium setting to interpret the magnetic response of CuGeO$_3$ to selective phonon driving, and possibly to modulate the spin-Peierls transition itself.

\begin{figure}[t]
\includegraphics[width=\columnwidth]{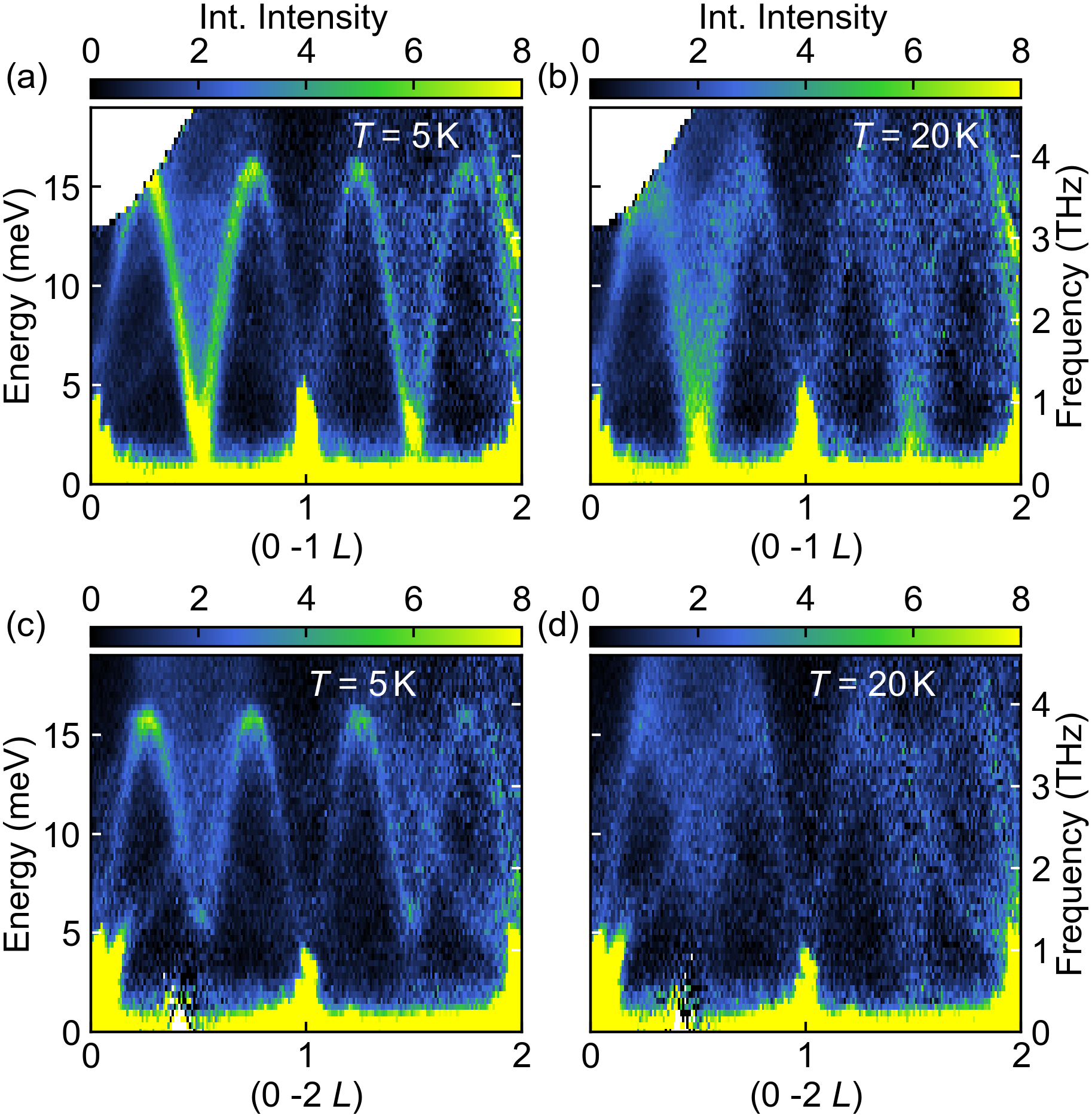}
\caption{Magnetic excitation spectrum of CuGeO$_3$, measured at 5~K with neutrons of incident energy $E_{\rm i} = 35.0$~meV.
(a,b) Chain direction ($L$) for $K = -1$ at $T = 5$ and $20$~K with integration ranges $-1.75 < H < 1.75$ and $-0.85 < K < -1.15$. (c,d) Chain direction ($L$) for $K = -2$ at $T = 5$ and 20~K with integration ranges $-1.75 < H < 1.75$ and $-1.85 < K < -2.15$.}
\label{fig:spinspectrum}
\end{figure}

\subsection{Mutual effects of phonons and magnetic excitations}

\begin{figure}[t]
\includegraphics[width=\linewidth]{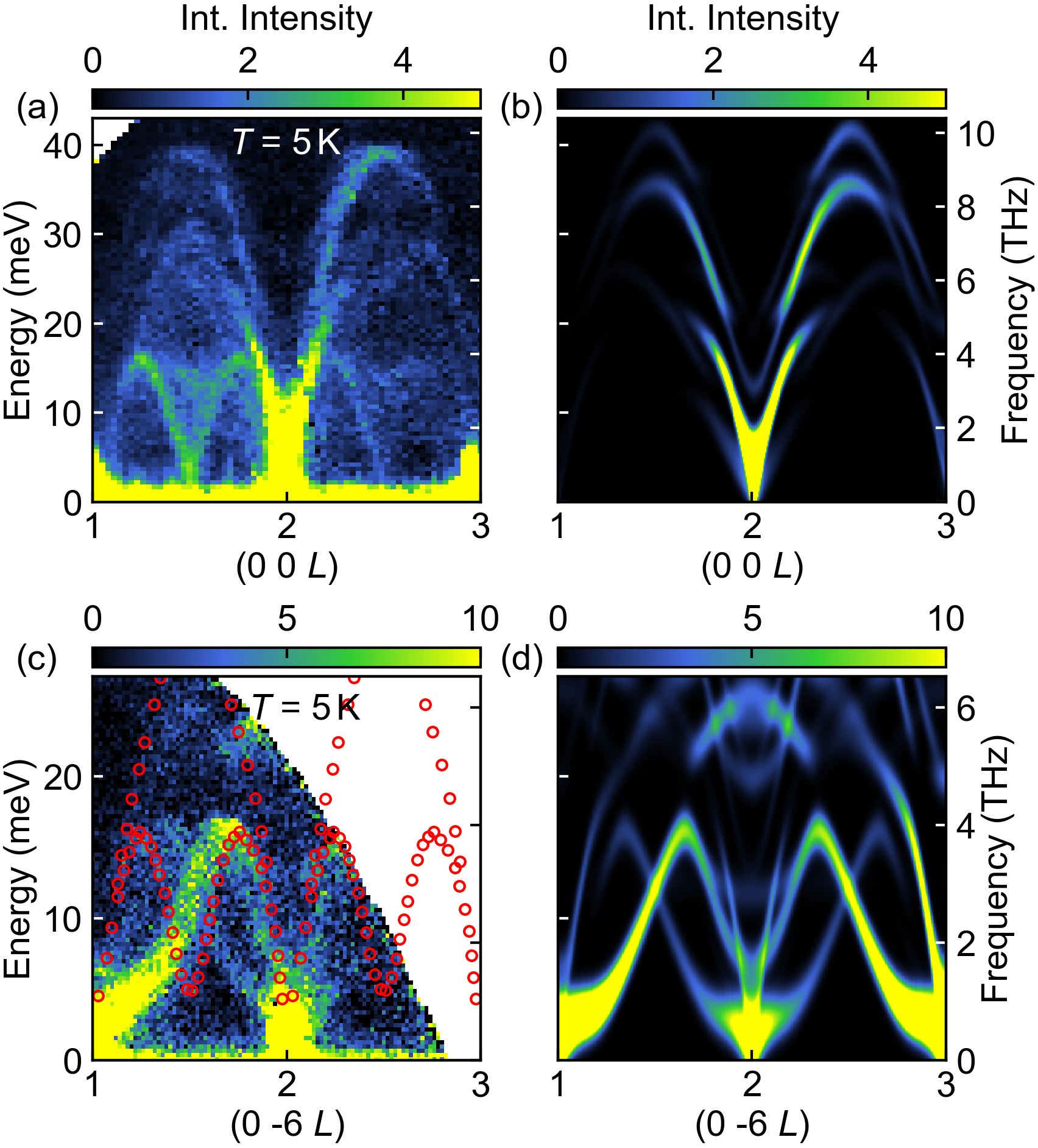}
\caption{Specific phonon spectra showing complex effects at energies and wave vectors related to the magnetic spectrum. (a,b) Dispersion along $L$ at $H = K = 0$, showing the INS and DFT spectra of the $B_{3u}$ acoustic phonon and related modes of the same symmetry, whose dispersions track the upper boundary of the spinon continuum. The INS data were measured with $E_{\rm i} = 54.4$~meV and the integration ranges are $-0.3 < H < 0.3$ and $-1 < K < 1$. (c,d) Dispersion along $L$ at $H = 0$ and $K = -6$, showing the INS and DFT spectra of the $B_{3u}$ acoustic phonon, whose maximum lies almost exactly at the maximum of the triplon dispersion. The INS data were measured with $E_{\rm i} = 35.0$~meV and the integration ranges are $-6.2 < K < -5.8$ and $-0.2 < H < 0.2$. 
Red dots indicate the lower and upper boundaries of the spin excitation spectrum.}
\label{fig:spinphonons}
\end{figure}

As a benchmark for detecting the influence of spin-phonon hybridization, we first present the magnetic spectrum. All information concerning our experimental measurements and conditions is provided in Sec.~\ref{sec:mm}D. The three measurement temperatures were chosen deep in the spin-Peierls phase (5~K), just above the spin-Peierls transition (20~K), and around the assumed limit of any magnetic correlation effects at $T \approx J_1$ (specifically, 110~K). To focus on the triplon mode of the dimerized phase, in Fig.~\ref{fig:spinspectrum} we show the spectrum measured near the Brillouin zone center with incident energy $E_{\rm i} = 35.0$~meV. At the lowest temperature [Figs.~\ref{fig:spinspectrum}(a,c)] we observe the triplon band minimum at 2 meV, the saddle point at the Brillouin-zone corner at 6 meV, and the upper band edge around 16 meV. These results are fully consistent with the early measurements of the triplon dispersion \cite{nishi94,regna96a,arai96}, which formed the basis for the most detailed fits performed to extract the magnetic interaction parameters \cite{uhrig97a,knett01}, and also with measurements of the full spinon continuum \cite{arai96,ikeuc13} [represented in Figs.~\ref{fig:scheme}(d,e) and visible in our Fig.~\ref{fig:spinphonons}]. 

At a temperature above $T_{SP}$, we find distinctive changes to the magnetic spectrum, as shown in Figs.~\ref{fig:spinspectrum}(b,d). The gap closes at the half-integer $L$ points and significant spectral weight is transferred out of the triplon branch to the spinon continuum. However, there are no places in the magnetic excitation spectrum where we observe any obvious disruption due to their hybridization with a phonon mode, or at least not of the conventional avoided-crossing form we studied for phonon-phonon hybridization in Sec.~\ref{sec:comp}B. One possible reason for this is that the phonon-triplon interaction vertex in a purely Heisenberg system such as CuGeO$_3$, which has the form $\sum_i g (b_i + b_i^\dag) \, \vec{S}_{1,i} \! \cdot \! \vec{S}_{2,i}$, mixes one phonon ($b_i$) with pairs of triplons rather than with a single triplon. 

Similarly, one may inspect the phonon spectrum for evidence of alterations arising from the magnetic sector. Once again, magnetic effects that would give rise to features in the measured phonon spectra not captured by our DFT results are extremely hard to find, even when the phonon modes are located at precisely the same places in energy and wave vector as the primary features of the magnetic spectrum. In Figs.~\ref{fig:spinphonons}(a,b) we observe that the acoustic phonon in the $L$ direction, and the modes that form its effective continuation beyond the avoided crossings, in essence wrap around the entire spinon continuum, peaking at an energy of 40 meV. In Figs.~\ref{fig:spinphonons}(c,d) we show that, far from the Brillouin-zone center where the phonon intensities are much stronger, one may observe the same acoustic phonon peaking at almost exactly the same energy as, and with just a small wave-vector offset from, the primary triplon branch. However, the lack of mixing features indicates that these are simply coincidences, presumably arising from the very large number of phonon modes. We stress here that we are not excluding any kind of spin-phonon coupling in CuGeO$_3$, given that the material has clear magnetoelastic phenomena in the form of a spin-Peierls transition. Rather, our statement is that we measure phonons already renormalized by magnetic effects (captured within our DFT calculations) and triplons already renormalized by lattice effects, which together provide an excellent account of the excitation spectra without leaving further anomalies requiring a more specific explanation. 

We close by remarking that this picture of mutual renormalization is in essence static, with the static structure also dictating the spectra of single phonon and triplon excitations. However, as noted in Sec.~\ref{sintro}, ultrafast laser technology in the THz domain has enabled previously unavailable types of dynamical and nonequilibrium experiment and analysis. On the technical side is the field of 2D coherent spectroscopy (2DCS, or more generally ``multidimensional" CS) \cite{liu25}, where two (or more) incident pulses are used to gain access to two time dimensions and thereby to separate the components of composite hybrid excitations. On the physics side, driving phonons with ultrafast coherent THz light opens up the phenomena of nonlinear phononics \cite{foers11,subed14,juras17} and magnetophononics \cite{fechn18}. The simplest manifestation of nonlinear phononics in CuGeO$_3$ would be the population of phonon modes higher in the spectrum that lie close to the sum frequencies of driven low-frequency phonons, for example populating 6 THz phonons (the yellow group in Table~\ref{tab:w}) by the combination of 3 THz phonons (red group). In particular, this mechanism forms an important route for the nonequilibrium population of coherent Raman-active phonons. Magnetophononics is a blanket term for the use of coherently driven THz phonons to control the magnetic properties, which can take place by a number of different pathways. Magnetophononic experiments have been performed on several ordered magnetic materials \cite{disa20,afana21,mashk21} and recently in a dimerized quantum magnet \cite{giorg23}. One anticipates that the results of Sec.~\ref{sec:sp}B will form a useful complement to the preliminary theoretical studies of magnetophononics in CuGeO$_3$ performed in Refs.~\cite{yarmo23,demaz25}. 

\section{Discussion and conclusion}
\label{sec:dc}

We have performed high-precision DFT calculations of the phonon spectrum of CuGeO$_3$ in its dimerized phase. The stability of our ground state is demonstrated by the absence of negative-energy modes and the accuracy of our spectra by the quantitative comparison with existing experimental data for the undimerized phase. To effect a quantitative comparison with the dimerized phase, we have performed state-of-the-art inelastic neutron scattering measurements of the phonon spectrum using a large sample and a modern spectrometer that allowed access to multiple Brillouin zones and a hierarchy of energy resolutions, providing unprecedented insight into the full spectral function of CuGeO$_3$. The unique level of comparison between theory and experiment confirms both the quality of the modelling and the intrinsic nature of the observed spectral features. 

Our DFT results contain full information about the phonon dispersions, intensities, and eigenvectors. The eigenvectors are particularly rich in physical information, providing a complete and consistent understanding of how different types of atomic motion give rise to different symmetries and energies as a result of the effective ``spring stiffnesses'' for different ions and lattice directions. Their {\bf Q}-dependence explains the wealth of phonon crossings and anticrossings throughout the spectrum. This complete 3D information forms a systematic basis for planning the geometry and polarization of coherent THz pumping experiments using CuGeO$_3$, which is a promising candidate material for both nonlinear phononics and magnetophononic phenomena.

Concerning DFT, we have shown that state-of-the-art methods for handling strongly correlated electronic and magnetic systems can provide quantitatively accurate phonon spectra. This is a significant achievement in a material as complex as CuGeO$_3$, where the lattice structure is so sensitive to the magnetic interactions that it undergoes a transition at only 14.2 K. Capturing this structure correctly is a precondition, which we have met in our calculations, but still some of the magnetic properties of CuGeO$_3$ remain as a challenge requiring further developments in DFT. This we find in trying to estimate the magnetic interactions, which have to be accurate to fractions of an meV in order to describe the near-gapless triplon dispersion, and measured against this challenging criterion our TEMA results are qualitative rather than quantitative.

With regard to CuGeO$_3$, we have taken another step towards its full quantitative characterization. Our equilibrium measurements confirm that the strong interaction between the lattice and magnetic sectors is largely manifest in establishing the spin-Peierls ground state. Beyond this, one observes rather well separated spectra arising from phonons and from spins, whose dressing by the other component is already included in their properties. Probing the extent of this mutual dressing, or renormalization, of the two components may be possible by out-of-equilibrium experimental methods.

In summary, we have computed and measured the phonon spectrum in the dimerized phase of the spin-Peierls material CuGeO$_3$. We characterize the energies, dispersions, and (anti)crossings of all 60 modes and relate these to the magnetic spectrum. We study the effects of these phonons on the spin interactions with a view to future ultrafast experiments probing the nature of magnetoelastic coupling in insulating quantum magnetic materials at a qualitatively deeper level of detail. 

\begin{acknowledgments}
We thank F. Ferrari and G. S. Uhrig for helpful discussions. AR, SB, and RV are grateful to the German Research Foundation (DFG) for financial support through projects TRR 288-422213477 (subproject A05 and B05) and VA 117/23-1--509751747. HL acknowledges support from the Royal Commission for the Exhibition of 1851. The work of MM was supported by the US Department of Energy, Office of Science, Basic Energy Sciences, under the award DE-SC-0018660. The neutron experiment at the Materials and Life Science Experimental Facility of the J-PARC was performed under a user program (Proposal No.~2023A0056).  

\end{acknowledgments}
	
\bibliography{phononbib}
	
\end{document}